\documentclass[conference]{IEEEtran}

\usepackage{cite}
\usepackage{amsmath,amssymb,amsfonts}
\usepackage{graphicx}
\usepackage{textcomp}
\usepackage{xcolor}
\usepackage{url}
\usepackage{hyperref}
\usepackage{booktabs}
\usepackage{listings}
\usepackage{multirow}
\usepackage{algorithm}
\usepackage{algpseudocode}
\usepackage{enumitem}
\usepackage{tikz}
\usetikzlibrary{arrows.meta, positioning, shapes, fit}

\title{QES-Backed Virtual FIDO2 Authenticators: Architectural Options for Secure, Synchronizable WebAuthn Credentials}

\author{
\IEEEauthorblockN{
Kemal Bicakci\IEEEauthorrefmark{1}\IEEEauthorrefmark{2},
Fatih Mehmet Varli\IEEEauthorrefmark{2},
Muhammet Emir Korkmaz\IEEEauthorrefmark{2},
Yusuf Uzunay\IEEEauthorrefmark{2}
}
\IEEEauthorblockA{\IEEEauthorrefmark{1}Informatics Institute, Istanbul Technical University, Ayazaga Campus, 34467, Maslak, Istanbul, TURKIYE}
\IEEEauthorblockA{\IEEEauthorrefmark{2}Securify Information Technology and Security Training Consulting Inc., 06378, Ankara, TURKIYE}
}

\begin{document}

\maketitle

\begin{abstract}
FIDO2 and the WebAuthn standard offer phishing-resistant, public-key based authentication but traditionally rely on device-bound cryptographic keys that are not naturally portable across user devices. Recent passkey deployments address this limitation by enabling multi-device credentials synchronized via platform-specific cloud ecosystems. However, these approaches require users and organizations to trust the corresponding cloud or phone providers with the protection and availability of their authentication material. In parallel, qualified electronic signature (QES) tokens and smart-card--based PKCS\#11 modules provide high-assurance, hardware-rooted identity, yet they are not directly compatible with WebAuthn flows.

This paper explores architectural options for bridging these technologies by securing a virtual FIDO2 authenticator with a QES-grade PKCS\#11 key and enabling encrypted cloud synchronization of FIDO2 private keys. We first present and implement a baseline architecture in which the cloud stores only ciphertext and the decryption capability remains anchored exclusively in the user's hardware token. We then propose a hardened variant that introduces an Oblivious Pseudorandom Function (OPRF)-based mechanism bound to a local user-verification factor, thereby mitigating cross-protocol misuse and ensuring that synchronization keys cannot be repurposed outside the intended FIDO2 semantics; this enhanced design is analyzed but not implemented. Both architectures preserve a pure WebAuthn/FIDO2 interface to relying parties while offering different trust and deployment trade-offs. We provide the system model, threat analysis, implementation of the baseline architecture, and experimental evaluation, followed by a discussion of the hardened variant's security implications for high-assurance authentication deployments.
\end{abstract}

\begin{IEEEkeywords}
FIDO2, WebAuthn, PKCS\#11, Qualified Electronic Signatures, Authentication, Cloud Synchronization, Virtual Authenticator, Identity Assurance
\end{IEEEkeywords}

% =============================================================
\section{Introduction}
% =============================================================

%The FIDO2 standard provides a robust, phishing-resistant authentication mechanism that eliminates the weaknesses associated with passwords. However, current FIDO2 authenticators store private keys in platform- or hardware-bound secure elements, limiting cross-device portability. In parallel, qualified electronic signature (QES) tokens, often exposed via PKCS\#11 interfaces, provide strong cryptographic identity guarantees but are not directly usable for WebAuthn.

%This paper bridges these two complementary technologies. We propose a novel client-side architecture where a Virtual FIDO2 Authenticator (VFA) is cryptographically protected and unlocked using a QES token. FIDO2 private keys are encrypted using a master key derived from the PKCS\#11 token and optionally stored on a cloud synchronization server.

%We make three main contributions:
%\begin{itemize}
%    \item We define a high-assurance hybrid authentication model combining FIDO2 virtual authenticators and QES tokens.
%    \item We design and implement a secure, PKCS\#11-anchored key management system enabling safe cloud-backed credential portability.
%    \item We evaluate the performance, usability, and security robustness of the system across realistic deployment scenarios.
%\end{itemize}

Passwordless authentication based on FIDO2 and the WebAuthn standard has gained significant momentum due to its strong resistance to phishing, credential reuse, and server-side compromise ~\cite{fido2,w3cwebauthn}. A long-standing limitation of FIDO2, however, is that authenticator private keys are typically bound to a single device or hardware token, complicating cross-device authentication and recovery. To improve usability, major ecosystem providers such as Google and Apple have recently introduced \emph{multi-device passkeys}, which synchronize WebAuthn credentials across user devices via cloud-backed keychain services~\cite{passkeysgoogle,applepasskeys}. While these passkey systems represent a major usability advancement, they implicitly require users and organizations to trust the corresponding cloud or mobile platform operator to securely store, manage, and synchronize authentication secrets.

In parallel, qualified electronic signature (QES) smart cards and PKCS\#11-compatible e-signature tokens remain the de facto standard for high-assurance identity verification in regulated and security-critical environments~\cite{pkcs11}. These devices provide hardware-based key protection and strong identity guarantees, but they are not natively compatible with WebAuthn flows and cannot directly serve as FIDO2 authenticators or passkey providers.

This paper addresses the gap between usability-oriented passkey synchronization and the assurance properties of hardware-secured identity tokens. We propose a hybrid client-side architecture in which a Virtual FIDO2 Authenticator (VFA) is cryptographically protected and locally unlocked using a PKCS\#11 e-signature token. In the baseline design, FIDO2 private keys are encrypted under a master secret derived from the token and may be safely synchronized across devices through an untrusted cloud service. Unlike platform passkeys, our approach ensures that the cloud provider never obtains access to sensitive keying material, preserving a strict hardware-root-of-trust model while still enabling multi-device credential portability.

In addition to the baseline design, we identify a class of cross-protocol misuse risks that may arise when cloud-synchronized key material is derived solely from token-based secrets. To mitigate these risks, we present an enhanced architectural option that incorporates an Oblivious Pseudorandom Function (OPRF)-based mechanism bound to a local user-verification factor. This hardened variant strengthens the derivation context and prevents synchronized keys from being repurposed outside the intended FIDO2 semantics. While the hardened design is analyzed at the architectural level, our prototype implementation focuses on the baseline model to evaluate feasibility, compatibility with existing browsers and WebAuthn servers, and performance considerations in realistic deployment scenarios.

Overall, our work demonstrates how a hardware-rooted trust model can be extended to support secure multi-device WebAuthn credential portability without relying on platform-controlled passkey ecosystems. We discuss the security implications, practical trade-offs, and integration paths for adopting such architectures in high-assurance authentication deployments.

\section{Related Work}

A broad range of prior work informs the design space of hardware-backed, multi-device authentication. Research on FIDO2 and WebAuthn has examined the security foundations and deployment challenges of passwordless credentials~\cite{fido2,w3cwebauthn,bonneau2012passwords,zhao2019webauthnformal,bicakci2022fido2}, while recent studies on platform passkeys and cloud-synchronized authenticators explore security, usability and cross-device availability trade-offs~\cite{passkeysgoogle,applepasskeys,lee2023passkey}. In parallel, the extensive literature on smart cards, qualified electronic signatures, and PKCS\#11-based key protection investigates high-assurance identity mechanisms~\cite{pkcs11,smartcardSurvey,etsiQES} and their integration into enterprise authentication systems. Finally, cryptographic techniques such as Oblivious Pseudorandom Functions (OPRFs)~\cite{oprf,opaque} and hardened key-derivation mechanisms~\cite{argon2} have been studied as methods for deriving secrets in a way that resists offline attacks and strengthens the overall security of authentication systems. In this section, we review these bodies of work and position our contribution at the intersection of secure credential portability, hardware-rooted authentication, and cloud-assisted synchronization.

\subsection{FIDO2 and WebAuthn Foundations}

The FIDO2 and WebAuthn standards~\cite{fido2,w3cwebauthn} have been extensively analyzed through empirical deployment studies~\cite{webauthndeployment}, threat modeling~\cite{kuchhal2023evaluating}, and formal verification~\cite{zhao2019webauthnformal}. Many earlier works highlight the security benefits of phishing-resistant authentication and the weaknesses of password-based systems~\cite{bonneau2012passwords}. However, they also emphasize limitations in credential portability and account recovery. Our work builds on these foundations by preserving the FIDO2 security model while addressing portability without relying on platform-operated passkey ecosystems.

\subsection{Multi-Device Credentials and Passkey Synchronization}

Recent industry deployments of multi-device passkeys~\cite{passkeysgoogle,applepasskeys} synchronize WebAuthn credentials through cloud-backed keychain systems. Academic studies examine security implications~\cite{lee2023passkey,bicakci2025balancing} and usability concerns~\cite{webauthndeployment}. These systems improve convenience but rely on trusting platform providers with the confidentiality and availability of synchronized secrets. In contrast, our approach maintains a hardware-rooted key hierarchy, preventing cloud access to any decryption capability.

\subsection{Smart Cards, QES Tokens, and PKCS\#11-Based Identity Systems}

Smart cards and QES-compliant signature tokens represent the primary mechanism for high-assurance identity in regulated environments~\cite{etsiQES,smartcardSurvey}. PKCS\#11~\cite{pkcs11} defines widely-used interfaces for hardware-backed key management and certificate operations. Prior work examines their security guarantees~\cite{pkcs11security} and integration into enterprise identity systems~\cite{pivcac}. However, these devices do not natively implement WebAuthn or CTAP2. Our work bridges this gap by anchoring a Virtual FIDO2 Authenticator to a PKCS\#11-protected identity module.

\subsection{Hybrid Authentication Models}

A number of systems aim to combine traditional PKI-based identity mechanisms with modern passwordless authentication frameworks. Enterprise deployments frequently integrate PIV/CAC smart cards with WebAuthn or federated login systems to achieve higher assurance levels, and several commercial solutions incorporate certificate-backed device identities into passwordless workflows. Recent academic work has explored deeper forms of integration. For example, FeIDo~\cite{feido} proposes a virtual FIDO2 token that derives WebAuthn credentials from stable, personally identifying attributes extracted from electronic identity documents (eIDs), such as passports. By binding FIDO2 credentials to long-lived eID attributes and protecting them within hardware-secured execution environments (e.g., SGX enclaves), FeIDo enables recovery from token loss and supports verifiable yet privacy-preserving meta-attributes for use cases such as anonymous age verification.

These hybrid systems demonstrate that PKI-based identities and FIDO2 authenticators can be combined to address shortcomings of hardware tokens, including recoverability and portability. However, existing models either rely on trusted platform components, do not support secure multi-device synchronization of WebAuthn private keys, or do not anchor credential protection in external hardware security modules such as QES or PKCS\#11 tokens. Our work differs by using a PKCS\#11-backed identity token as a stable hardware root of trust for a Virtual FIDO2 Authenticator and by enabling cloud-assisted synchronization of encrypted credentials under a strictly minimized trust model.

\subsection{Virtual and Software-Based FIDO2 Authenticators}

Software-based or "virtual" authenticators appear in browser environments and testing frameworks, such as Chrome's Virtual Authenticator API~\cite{chromevirtualauth}, WebDriver-based CTAP2 emulators~\cite{webdriverctap}, and open-source implementations like SoftFIDO and virtual-ctap2~\cite{softfido}. These systems demonstrate the viability of software authenticators but do not include secure key protection or cross-device synchronization. Our work extends this model by introducing a Virtual FIDO2 Authenticator protected by a PKCS\#11-backed master key.

\subsection{OPRFs and Hardened Key-Derivation Mechanisms}

OPRFs~\cite{oprf} and OPRF-based protocols such as OPAQUE~\cite{opaque} provide privacy-preserving key derivation and resistance to credential misuse. Related work explores augmented PAKE constructions~\cite{PAKE} and memory-hard KDFs like Argon2~\cite{argon2}. We leverage these mechanisms not for password authentication but for strengthening the derivation of synchronized keys, binding them to a local user-verification factor and preventing cross-protocol repurposing.

% =============================================================
\section{Problem Definition}
\label{sec:problem-definition}
% =============================================================

Despite the usability and security improvements introduced by FIDO2, several limitations remain for environments requiring both high assurance and multi-device portability. In this section, we formalize the problem and articulate the design requirements that motivate our hybrid architecture.

\subsection{Limitations of Device-Bound FIDO2 Credentials}
Conventional FIDO2 authenticators---whether platform authenticators (e.g., Secure Enclave, Android Keystore, Windows Hello) or roaming hardware keys---store private keys in device-local secure hardware. While effective for protecting against extraction, this device binding hinders portability: users must register each authenticator separately on every service and must repeat registration if a device is lost or replaced. This model does not naturally support cross-device roaming or backup without introducing additional trust assumptions.

\subsection{Trust Assumptions in Passkey Synchronization}
Recent passkey ecosystems achieve multi-device capability through cloud‐backed credential synchronization. Although cryptographic protections are employed, users and organizations must inherently trust the supporting cloud provider (e.g., Apple, Google, Microsoft) to correctly implement synchronization, protect encrypted material, and maintain availability. In certain security-sensitive or regulated settings—e.g., public-sector identity, financial services, critical infrastructure—relying on consumer cloud providers introduces organizational risk and governance challenges.

\subsection{Incompatibility of PKCS\#11 Tokens with WebAuthn}
In contrast, PKCS\#11 smart cards and qualified electronic signature (QES) tokens provide high-assurance, hardware-rooted key protection with strong identity guarantees. However, these devices are not compatible with WebAuthn's CTAP2 protocol and cannot operate as native FIDO2 authenticators. As a result, organizations deploying both WebAuthn and QES systems must maintain parallel authentication infrastructures, increasing operational complexity.

\subsection{Need for High-Assurance, Cloud-Portable WebAuthn Credentials}

The resulting research gap can be characterized by the need for an authentication architecture that:
\begin{itemize}
    \item preserves a pure WebAuthn/FIDO2 model on the server side,
    \item enables cross-device credential portability without entrusting authentication secrets to consumer cloud providers,
    \item leverages an existing high-assurance PKCS\#11 identity token as the cryptographic root of trust, and
    \item treats the cloud service used for synchronization as entirely untrusted with respect to the confidentiality and integrity of key material.
\end{itemize}

\subsection{Threat Model}
We consider the following classes of adversaries:
\begin{itemize}
    \item \textbf{Remote malware}: may run arbitrary code on the client device and access local storage, but cannot extract sensitive data from protected process memory.
    \item \textbf{Cloud adversary}: gains full access to all synchronized ciphertext stored on the cloud service.
    \item \textbf{Physical attacker}: can steal the client device and/or the PKCS\#11 token.
    \item \textbf{Network attacker}: can observe, block, or modify traffic between client, cloud, and relying party.
\end{itemize}

We assume that the PKCS\#11 token enforces local PIN verification, prevents private-key extraction, and provides physical tamper resistance consistent with its certification level.

Under these assumptions, our goal is to ensure the confidentiality of FIDO2 private keys, prevent unauthorized use of virtual authenticator credentials, and enable secure multi-device recovery without requiring any modifications to WebAuthn relying parties.

% =============================================================
\section{Proposed Solution}
\label{sec:solution}
% =============================================================

\begin{figure*}[t]
    \centering
\begin{tikzpicture}[
  node distance=0.8cm and 1.5cm,
  box/.style = {draw, rounded corners, thick, align=center, inner sep=6pt},
  smallbox/.style = {draw, rounded corners, align=center, inner sep=4pt},
  arrow/.style = {->, >=Stealth, thick}
]

% Nodes
\node[box, fill=gray!10] (user) {User};

\node[box, fill=blue!5, below=of user] (browser) {Web Browser\\(WebAuthn client)};

\node[box, fill=orange!10, below=of browser] (os) {OS CTAP Interface};

\node[box, fill=green!10, below left=1.5cm and -0.4cm of os] (vfa) {Virtual FIDO2\\Authenticator App};

\node[box, fill=yellow!20, below=of vfa] (pkcs11) {PKCS\#11\\QES / Smart Card\\E-Signature Token};

\node[box, fill=purple!10, below right=1.5cm and -0.4cm of os] (cloud) {Cloud Credential Store\\(Encrypted FIDO Keys)};

\node[box, fill=red!5, right=3.6cm of browser] (server) {Relying Party Server\\(Pure WebAuthn)};

% Arrows
\draw[arrow] (user) -- (browser) node[midway,left] {\footnotesize UI};
\draw[arrow] (browser) -- (os) node[midway,left] {\footnotesize WebAuthn};
\draw[arrow] (os) -- (vfa) node[midway,left] {\footnotesize CTAP2 / HID};
\draw[arrow] (vfa) -- (pkcs11) node[midway,left] {\footnotesize PKCS\#11 API};
%\draw[arrow] (vfa) -- (cloud) node[midway,right] {\footnotesize Enc/Dec \& Sync};
\draw[arrow] (vfa) -- (cloud)
    node[midway, above] {\footnotesize Enc/Dec \& Sync};

\draw[arrow] (browser) -- (server) node[midway,above] {\footnotesize HTTPS / WebAuthn};
%\draw[arrow] (server) -- (browser) node[midway,below] {\footnotesize Challenges};

% Braces / annotations
%\node[smallbox, fill=white, draw=none, above right=0.1cm and -0.5cm of vfa] (vfaLabel) {\footnotesize Derive $K_{\text{master}}$ from token};

\node[smallbox, fill=white, draw=none, right=0.1cm of pkcs11] {\footnotesize Non-exportable key};

\node[smallbox, fill=white, draw=none, above=0.05cm of cloud] {\footnotesize Ciphertexts only};

\end{tikzpicture}
\caption{High-level architecture: the virtual FIDO2 authenticator is locally protected by a PKCS\#11 e-signature token and uses encrypted cloud storage for synchronizable FIDO credentials. The server remains a pure WebAuthn/FIDO2 relying party.}
\label{fig:architecture}
\end{figure*}
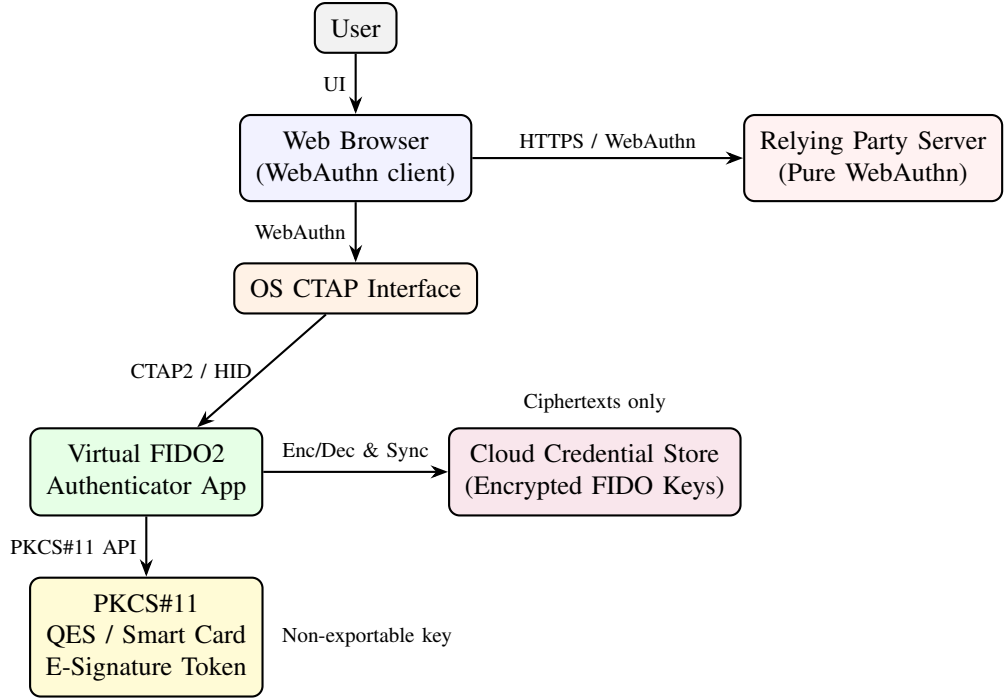

We propose a hybrid authentication architecture that enables secure, multi-device FIDO2 credential portability while preserving a hardware-rooted trust model. Our approach centers around a Virtual FIDO2 Authenticator (VFA) that is cryptographically bound to a PKCS\#11 e-signature token and synchronized through an untrusted cloud service.

\subsection{System Overview}

The architecture consists of two primary client-side components and a single untrusted cloud repository (Fig.~\ref{fig:architecture}):

\begin{enumerate}[label=\Roman*.]
    \item \textbf{Virtual FIDO2 Authenticator (VFA)}: a software authenticator implementing CTAP2 semantics and exposing itself to the browser by emulating a USB Human Interface Device (HID), as a replacement for a hardware FIDO2 security key. It also incorporates a local and cloud-synchronized repository containing ciphertexts of FIDO2 private keys and associated metadata.
    \item \textbf{PKCS\#11 E-Signature Token}: a hardware token storing a high-assurance private key that is used to derive the master encryption key via deterministic signing (or, if supported by the token’s capabilities, unwrap a previously wrapped master key for protecting FIDO2 material).    
    %\item \textbf{Encrypted Credential Store}: a local and cloud-synchronized repository containing ciphertexts of FIDO2 private keys and associated metadata.
    \item \textbf{Untrusted Cloud Sync Server}: holds only encrypted blobs; never receives cleartext key material.
\end{enumerate}

The VFA handles WebAuthn \texttt{MakeCredential} and \texttt{GetAssertion} operations normally, but stores all corresponding credential private keys encrypted under a master key that is anchored to the PKCS\#11 token.

\subsection{Master Key Derivation or Unwrapping}

We consider two mechanisms depending on token capabilities:

\subsubsection{Deterministic Derivation from PKCS\#11 Signatures}

If the PKCS\#11 private key supports deterministic signatures (e.g., RSA PKCS\#1~v1.5 or deterministic EdDSA), we compute:

\begin{equation}
\sigma = \mathrm{Sign}_P\!\left(H(\text{label})\right)
\label{eq:sigma}
\end{equation}

\begin{equation}
K_{\mathrm{master}} = \mathrm{HKDF}\!\left(\sigma, \text{``VFA-MK''}\right)
\label{eq:two}
\end{equation}

where the label is a fixed string and $\sigma$ is stable across unlocks.

\subsubsection{Token-Assisted Unwrapping}

For non-deterministic algorithms (e.g., ECDSA with random $k$, RSA-PSS), the VFA generates a random symmetric $K_{\mathrm{master}}$ once during enrollment, encrypts it using a token-stored wrapping key, and stores only the encrypted blob. On each unlock, the token decrypts the blob, restoring $K_{\mathrm{master}}$ without ever exposing it outside a protected client session.

\subsection{Credential Encryption and Storage}

For each new FIDO2 credential, the VFA:

\begin{enumerate}
    \item generates the FIDO2 key pair $(k_\mathrm{FIDO}, K_\mathrm{FIDO})$,
    \item constructs metadata including RP ID, credential ID, user handle, and counters,
    \item encrypts the private key using $K_{\mathrm{master}}$:
   \begin{equation}
C = \mathrm{Enc}_{K_{\mathrm{master}}}\!\left(k_{\mathrm{FIDO}} \parallel \text{metadata}\right)
\end{equation}

    \item stores $C$ locally and synchronizes it to the cloud,
    \item returns standard WebAuthn attestation to the relying party.
\end{enumerate}

The cloud never receives unencrypted private keys, credential metadata, or $K_{\mathrm{master}}$.

\subsection{Unlock Procedure}

Before handling any \texttt{GetAssertion} request, the VFA must be unlocked:

\begin{algorithm}[h]
\caption{Unlock Procedure}
\begin{algorithmic}[1]
\State Prompt user to insert token and enter PIN
\If{derivation method}
    \State $\sigma \leftarrow \mathrm{Sign}_P(H(\text{label}))$
    \State $K_{\mathrm{master}} \leftarrow \mathrm{HKDF}(\sigma,\text{``VFA-MK''})$
\Else
    \State $K_{\mathrm{master}} \leftarrow \mathrm{DecryptWithToken}(C_{\mathrm{master}})$
\EndIf
\State Load encrypted credential store
\State Decrypt each credential using $K_{\mathrm{master}}$
\State Mark VFA as unlocked
\end{algorithmic}
\end{algorithm}

\subsection{Cloud Synchronization}

Synchronization is optional but enables multi-device functionality. The VFA periodically pushes ciphertext updates to the cloud and fetches new records when available. On a new device:

\begin{enumerate}
    \item the user installs the VFA,
    \item unlocks it using the same PKCS\#11 token,
    \item downloads the encrypted store,
    \item decrypts it using the derived or unwrapped $K_{\mathrm{master}}$,
    \item and imports the full set of FIDO2 credentials.
\end{enumerate}

No server-side WebAuthn logic requires modification.

\subsection{FIDO2 Runtime Operation}

When the browser issues a \texttt{MakeCredential} or \texttt{GetAssertion} request through the CTAP interface, the VFA responds identically to a hardware authenticator:

\begin{itemize}
    \item counters and flags are updated in software,
    \item signature operations use the decrypted FIDO2 private key,
    \item authenticatorData and clientDataHash structures conform to the WebAuthn specification.
\end{itemize}

This ensures full compatibility with existing FIDO2 relying parties without requiring attestation changes or server-side extensions.

\section{Performance Evaluation}

We evaluate the performance impact of introducing a PKCS\#11-bound Virtual FIDO2 Authenticator (VFA), focusing on the costs of master-key derivation, credential operations, and cloud synchronization. Measurements were conducted on a typical workstation (Intel i7 CPU, 16\,GB RAM, a standard USB e-signature token accessed via PKCS\#11, and a commodity cloud object store), and averaged over 1{,}000 runs. The results demonstrate that the proposed architecture introduces moderate overhead compared to hardware FIDO2 authenticators, yet remains well within the latency bounds acceptable for interactive authentication.

\subsection{PKCS\#11 Operation Latency}

In our implementation, the VFA derives the master key exclusively through \emph{deterministic signing} using the PKCS\#11 private key. Each unlock operation computes $\sigma$ as defined in eq.(\ref{eq:sigma}), which is then used as input to HKDF to recover $K_{\mathrm{master}}$ as in eq.(\ref{eq:two}). Signing operations on the USB e-signature token therefore constitute the main performance cost during the unlock phase. Our measurements indicate an average latency of 42\,ms per signing operation, consistent with typical commercial USB-based qualified e-signature tokens. Since this step is performed only during unlocking—and not for each authentication request—its impact on usability is limited.

\subsection{FIDO2 Credential Operations}

Once unlocked, the VFA performs all FIDO2 operations in software. We measured the execution time of \texttt{MakeCredential} (credential registration) and \texttt{GetAssertion} (authentication). \texttt{MakeCredential} requires generating a new key pair and producing an attestation signature, whereas \texttt{GetAssertion} requires a single assertion signature. Measured mean latencies of 15\,ms and 7\,ms respectively indicate that the VFA remains efficient for interactive authentication, comparable to other software-based authenticators.

\subsection{Cloud Synchronization Overhead}

To support multi-device portability, the VFA synchronizes encrypted credential material with an untrusted cloud repository. Synchronization occurs asynchronously and outside the critical login path. Downloading the encrypted credential store required on average 220\,ms, primarily due to network and cloud storage latency. Upload performance was similar. Because synchronization happens only during onboarding or incremental updates, it does not hinder regular authentication workflows.

\subsection{Comparison with Hardware and Platform Authenticators}

Hardware FIDO2 keys generally exhibit \texttt{GetAssertion} latencies between 5--10\,ms and \texttt{MakeCredential} latencies of 10--25\,ms, depending on the attestation mode and cryptographic hardware. Platform authenticators (e.g., Secure Enclave or TPM-backed implementations) may achieve slightly lower values due to hardware acceleration. The performance of the VFA fits within the expectations for high-level software authenticators, with deterministic PKCS\#11 signing introducing only a one-time overhead during unlock.

\subsection{Summary of Results}

Table~\ref{tab:latency} summarizes the measured operation latencies. The cost introduced by PKCS\#11-based master-key derivation occurs only during unlocking, while normal FIDO2 operations remain lightweight. Overall, the system provides responsive authentication performance suitable for everyday use, even when cloud synchronization is enabled.

\begin{table}[h]
\centering
\caption{Latency Comparison of Operations}
\label{tab:latency}
\begin{tabular}{lcc}
\toprule
Operation & Mean (ms) & Std Dev \\
\midrule
PKCS\#11 Sign (Deterministic) & 42 & 5.2 \\
FIDO2 MakeCredential & 15 & 1.1 \\
FIDO2 GetAssertion & 7 & 0.9 \\
Cloud Sync Download & 220 & 21 \\
\bottomrule
\end{tabular}
\end{table}

% =============================================================
\section{Security Analysis}
% =============================================================

We now analyze the security properties of the proposed architecture under the threat model defined earlier. Our analysis focuses on key confidentiality, resistance to device and cloud compromise, comparison with existing authenticator models, and inherent limitations arising from the WebAuthn security model. Fig.~\ref{fig:threat-model} summarizes the adversarial interactions and the security goals that the system aims to guarantee.

\subsection{Threat Model}

\begin{figure*}[t]
\centering
\resizebox{\textwidth}{!}{%
\begin{tikzpicture}[
  node distance=0.9cm and 2.0cm,
  entity/.style   = {draw, rounded corners, thick, align=center,
                     inner sep=5pt, fill=gray!5},
  attacker/.style = {draw, rounded corners, thick, align=center,
                     inner sep=5pt, fill=red!10},
  goals/.style    = {draw, rounded corners, thick, align=left,
                     inner sep=6pt, fill=green!10},
  arrow/.style    = {->, >=Stealth, thick},
  dashedarrow/.style = {->, >=Stealth, thick, dashed}
]

% --- Honest entities (center column) ---
\node[entity] (user)  {User};

\node[entity, below=of user] (client) {Client Device\\(Browser + OS + VFA)};

\node[entity, below=of client] (token) {PKCS\#11\\E-Sign Token};

\node[entity, below=of token] (cloud) {Cloud Store\\(Encrypted FIDO Keys)};

\node[entity, right=4.0cm of client] (server) {Relying Party\\(WebAuthn Server)};

% --- Attackers ---
\node[attacker, left=4.0cm of client] (remoteMalware) {Remote Malware};

\node[attacker, left=3.8cm of cloud] (cloudAdv) {Cloud Adversary};

\node[attacker, right=4.0cm of token] (physicalAdv) {Physical Attacker};

% --- Honest edges ---
\draw[arrow] (user) -- (client);

\draw[arrow] (client) -- node[right,pos=0.5] {\footnotesize PKCS\#11} (token);

\draw[arrow] (token) -- node[right,pos=0.5] {\footnotesize Sync} (cloud);

\draw[arrow] (client) -- node[above,pos=0.5] {\footnotesize WebAuthn/TLS} (server);

% --- Attack edges ---
\draw[dashedarrow] (remoteMalware) -- node[above,pos=0.5] {\footnotesize Code Injection} (client);

\draw[dashedarrow] (cloudAdv) -- node[above,pos=0.5] {\footnotesize DB Compromise} (cloud);

\draw[dashedarrow] (physicalAdv) -- node[above,pos=0.5] {\footnotesize Token Theft} (token);

\draw[dashedarrow] (physicalAdv) -- node[below,pos=0.5, xshift=-12pt] {\footnotesize Device Theft} (client);

% --- Security goals box ---
\node[goals, below=1.3cm of cloud] (goals) {%
\footnotesize
\textbf{Security Goals:}\\
1) Cloud compromise $\Rightarrow$ no key disclosure.\\
2) Device theft $\Rightarrow$ no unlock without token.\\
3) Token theft $\Rightarrow$ protected by PIN/PUK.\\
4) Malware without token $\Rightarrow$ no FIDO use.
};

\end{tikzpicture}
}
\caption{Threat model for the proposed architecture. Adversaries include remote malware, a cloud attacker with database access, and a physical attacker capable of stealing the device and/or the e-signature token. The design aims to preserve FIDO2 private-key confidentiality and prevent unauthorized authentications without the PKCS\#11 token.}
\label{fig:threat-model}
\end{figure*}
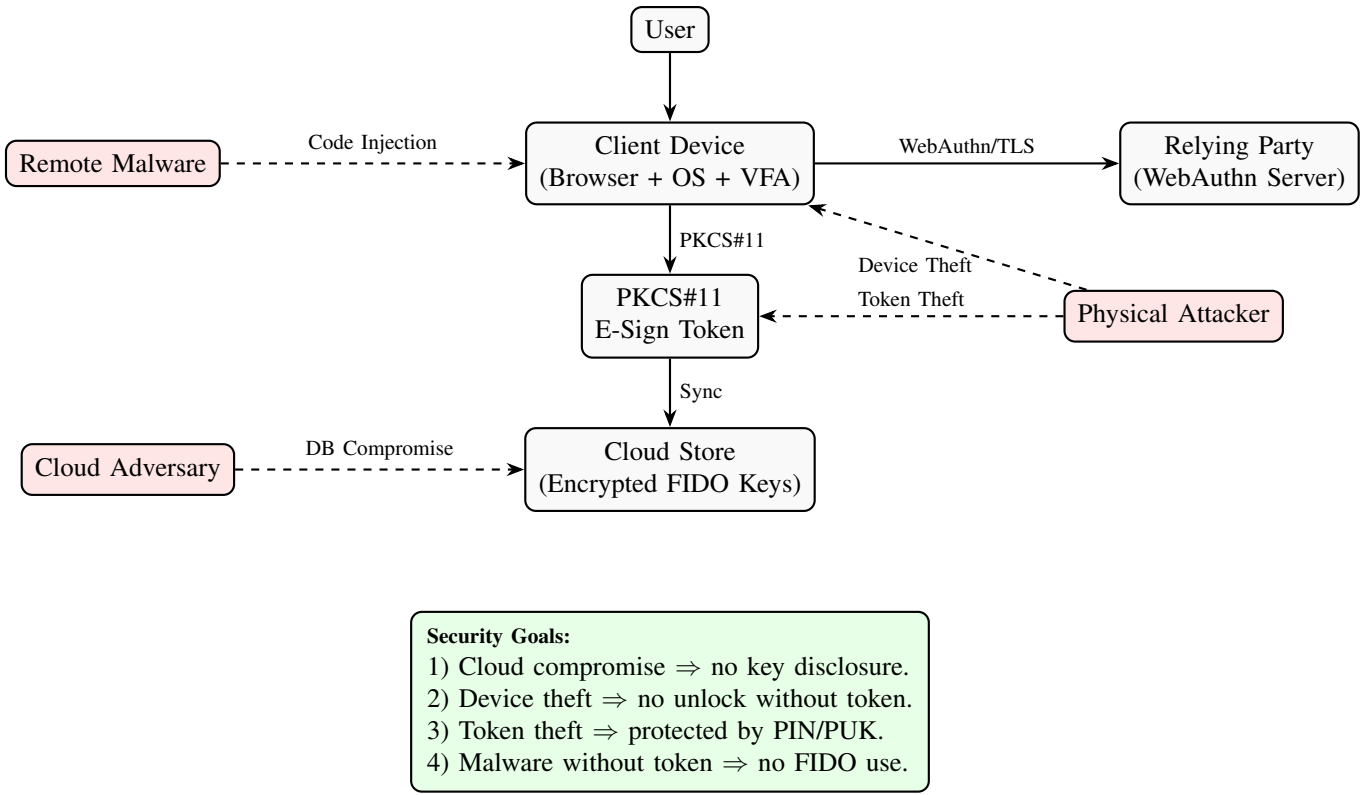

We now analyze the security properties of the proposed architecture under the threat model defined in Section~\ref{sec:problem-definition} and illustrated in Fig.~\ref{fig:threat-model}.

\subsection{Key Protection Guarantees}

A central design goal is to ensure that FIDO2 private keys remain confidential even if the client device or synchronization infrastructure is compromised. In the proposed architecture, each FIDO2 private key $k_{\mathrm{FIDO}}$ and its associated metadata are encrypted under the master key $K_{\mathrm{master}}$ derived from the PKCS\#11 token, yielding the ciphertext. Neither the encrypted credential store on the client nor the synchronized data in the cloud contains $K_{\mathrm{master}}$. The master key is recovered only during the unlock procedure, where the token performs deterministic signing as in (\ref{eq:sigma}) or decrypts a wrapped key, and is held in protected process memory for the duration of the unlocked session.

Under the assumption that the adversary cannot read sensitive material from protected process memory, a compromise of the client filesystem or cloud storage reveals only ciphertexts. Without the PKCS\#11 token and its PIN, the attacker cannot recover $K_{\mathrm{master}}$ and thus cannot decrypt any FIDO2 private keys. This prevents credential cloning and exfiltration, even under full compromise of the cloud synchronization service.

\subsection{Resistance to Device Theft}

If an attacker steals the client device but not the PKCS\#11 token, the encrypted credential store remains unusable. Local disk access alone does not suffice to recover $K_{\mathrm{master}}$, as it is never stored in cleartext and can only be reconstructed by interacting with the token. As a result, an attacker cannot impersonate the user at WebAuthn relying parties, nor can they migrate FIDO2 credentials to another device.

If both the device and token are stolen, the security reduces to that of the PKCS\#11 token itself. In our model, the token enforces local PIN or PUK entry and is designed to resist offline key extraction. The attacker must therefore guess or obtain the PIN to unlock the token; until then, $K_{\mathrm{master}}$ and the underlying FIDO2 private keys remain protected.

\subsection{Server-Side Independence}

From the perspective of the relying party, the VFA behaves identically to a standard FIDO2 authenticator. The server implements an unmodified WebAuthn interface, verifies attestation and assertion signatures as usual, and stores only public keys and credential identifiers. No PKCS\#11-specific logic, master-key management, or cloud synchronization awareness is required on the server side. This reduces the attack surface of the relying party, simplifies deployment, and ensures compatibility with existing FIDO2 infrastructure and libraries.

\subsection{Comparison to Hardware FIDO Keys and Passkeys}

Compared to hardware FIDO2 tokens (e.g., USB security keys), the proposed architecture offers similar protection against key extraction: in both cases, an attacker cannot obtain long-term signing keys for use on another device. Hardware tokens typically perform all signing operations within tamper-resistant hardware, whereas the VFA executes FIDO2 signatures in software after deriving $K_{\mathrm{master}}$. However, in our design, private keys are decrypted only into protected process memory and are never stored in cleartext at rest or in the cloud.

Relative to platform passkeys synchronized by ecosystem providers (e.g., Apple, Google, Microsoft), our architecture adopts a more constrained trust model. Passkeys rely on vendor-managed keychain services and cloud infrastructure, whereas our design treats the cloud as untrusted and anchors all decryption capability in an external PKCS\#11 token. This improves privacy and governance in environments where trusting consumer cloud providers is undesirable.

% =============================================================
\subsection{Design Choice: Deterministic Credential Derivation vs. Encrypted Credential Storage}
\label{sec:design-choice}
% =============================================================

A fundamental design choice for synchronizable WebAuthn credentials is whether each relying-party (RP) private key should be
(i) deterministically derived from a strong long-term secret, or
(ii) generated randomly and protected using a master encryption key anchored in hardware.
In this work, we adopt the second approach.

\paragraph{Option 1 - Deterministic derivation from $K_{\mathrm{master}}$}
A stateless construction derives a per-RP credential secret as
\begin{equation}
  k_{\mathrm{FIDO}}^{rp} = \mathrm{KDF}(\mathrm{rpid}, K_{\mathrm{master}})
\end{equation}

where $\mathrm{rpid}$ is the RP identifier and $K_{\mathrm{master}}$ is a strong secret recoverable across devices.
This approach simplifies backup and recovery, as possession of $K_{\mathrm{master}}$ suffices to reconstruct all credentials.
However, it introduces several practical limitations in real WebAuthn deployments.
First, WebAuthn commonly requires supporting \emph{multiple credentials per RP} (e.g., multiple accounts, resident credentials,
or re-registrations), which forces additional labels or state to avoid unintended key reuse.
Second, deterministic derivation is sensitive to RP identifier canonicalization; inconsistencies across platforms may lead to
credential loss or mismatches.
Third, credential lifecycle operations such as secure deletion and key rotation are difficult: derived keys remain recoverable
as long as $K_{\mathrm{master}}$ is valid, and rotating $K_{\mathrm{master}}$ invalidates all credentials simultaneously.

\paragraph{Option 2 - Random credential generation with encrypted storage under $K_{\mathrm{master}}$ (chosen)}
In the proposed architecture, the Virtual FIDO2 Authenticator (VFA) generates each WebAuthn credential keypair randomly and
stores the private key only in encrypted form:
\[
  C = \mathrm{Enc}_{K_{\mathrm{master}}}\big(k_{\mathrm{FIDO}} \parallel \text{metadata}\big)
\]
where $K_{\mathrm{master}}$ is a master encryption key recovered during unlock using the PKCS\#11 token, as described in
Section~\ref{sec:solution}. The master key is never stored in persistent storage and exists only transiently in protected
client memory after successful token-based user verification.

This design aligns naturally with the WebAuthn credential model. It supports multiple credentials per RP without additional
derivation labels, enables \emph{true deletion} by removing the corresponding encrypted record, and allows \emph{key rotation}
by re-encrypting stored credentials under a refreshed $K_{\mathrm{master}}$ without changing the underlying WebAuthn keys.
Moreover, requiring the presence of an encrypted credential record prevents a derive-on-demand oracle for arbitrary RP
identifiers.

\paragraph{Design Rationale}
Although both approaches can provide strong confidentiality when $K_{\mathrm{master}}$ is high-entropy and properly protected,
we select encrypted credential storage under $K_{\mathrm{master}}$ because it provides superior lifecycle management,
cleanly supports the WebAuthn data model, and avoids subtle derivation and canonicalization pitfalls.
The OPRF-based hardening discussed later strengthens the recovery and unlock path for $K_{\mathrm{master}}$ against
cross-protocol misuse, without altering the server-side WebAuthn interface.

\subsection{Binding FIDO2 Usage to QES Tokens for Strong Identity Assurance}
\label{sec:qes-binding}

In this work, \emph{WebAuthn compatibility} strictly refers to preserving an unmodified, standards-compliant WebAuthn/FIDO2
protocol execution between the browser and the relying party (RP), including unchanged registration and authentication
ceremonies, message formats, and server-side verification logic. Importantly, this notion of compatibility does not preclude
the RP from applying additional \emph{application-layer policies} during account enrollment or credential lifecycle
management.

Qualified electronic signature (QES) tokens provide strong, hardware-rooted identity guarantees, but WebAuthn does not natively
support expressing or verifying such identity semantics within its protocol messages. As a result, enforcing the use of a QES
token cannot be achieved inside the WebAuthn protocol itself without introducing non-standard extensions. Instead, strong
identity assurance can be imposed without breaking WebAuthn compatibility by positioning QES checks outside the protocol, at
the policy and enrollment layers.

\paragraph{Application-layer identity gating for enrollment}
A relying party may require a successful QES-based identity proof as a prerequisite to initiating the standard WebAuthn
\texttt{MakeCredential} ceremony. In this model, the user first performs a QES operation (e.g., signing a server-provided
challenge or presenting a certificate-backed authentication) that is verified by the RP using the qualified certificate
chain. Only upon successful verification does the RP authorize the WebAuthn registration, for example by issuing a
short-lived, session-bound registration token. The subsequent WebAuthn ceremony remains entirely standard and produces a
regular WebAuthn credential from the RP’s perspective. Thus, no WebAuthn messages, data structures, or verification logic are
modified, while credential creation is cryptographically tied to prior possession of a QES token.

\paragraph{Policy-based assurance during authenticator usage}
Beyond enrollment, the proposed architecture ensures that WebAuthn credentials cannot be used unless the Virtual FIDO2
Authenticator (VFA) is successfully unlocked using the PKCS\#11-backed QES token. All FIDO2 private keys remain encrypted under
a master key anchored to the token, and decryption is possible only after local token-based user verification (e.g., PIN
entry). This enforces a hardware-rooted usage constraint at the client side, independent of RP logic, while the RP continues
to validate only standard WebAuthn assertions.

Taken together, this separation of concerns preserves full WebAuthn interoperability while enabling deployments to mandate
high-assurance, hardware-rooted identity through QES tokens. The WebAuthn protocol remains unchanged and unaware of QES, while
identity assurance is enforced through a combination of enrollment-time policy decisions and client-side cryptographic
binding. This design avoids protocol-level modifications, maintains compatibility with existing browsers and servers, and
allows flexible adoption depending on regulatory and operational requirements.

\subsection{Cross-Protocol Attacks} \label{sec:cross-protocol}

The baseline architecture in Section~\ref{sec:solution} which derives the master key $K_{\mathrm{master}}$ directly from a PKCS\#11 signature on a fixed label, as shown in (\ref{eq:sigma})--(\ref{eq:two}), aims to bind all encrypted FIDO2 credentials to a non-exportable hardware key on the token. As long as an adversary cannot obtain the deterministic signature $\sigma = \mathrm{Sign}_P(H(\text{label}))$, they cannot reconstruct $K_{\mathrm{master}}$ and decrypt any synchronized credentials.

However, this construction introduces a specific class of \emph{cross-protocol} attacks. An adversary who is able to trick the user or the client device into invoking the same PKCS\#11 signing operation on $H(\text{label})$ \emph{outside} the VFA context can recover $\sigma$ and, consequently, $K_{\mathrm{master}}$ offline. Concretely, a malicious application, compromised middleware, or remote malware with access to the PKCS\#11 interface could present a signing request that appears unrelated to WebAuthn (e.g., as part of a document-signing or login flow), but internally uses the special input $H(\text{label})$. If the user approves this operation (e.g., by entering the PIN on a familiar token dialog), the adversary obtains the exact deterministic signature value that the VFA relies on for master-key derivation.

Once $\sigma$ is known, the attacker can locally compute
\[
K_{\mathrm{master}} = \mathrm{HKDF}\!\left(\sigma, \text{``VFA-MK''}\right)
\]
without further interaction with the token or the VFA. At that point, any ciphertexts obtained from the compromised device or from the cloud store become decryptable, breaking the intended hardware-root-of-trust guarantee for synchronized FIDO2 credentials. This attack does not require the token’s private key to be extracted; it merely exploits the fact that the same PKCS\#11 signing primitive is reused across protocol contexts with a recognizable input.

In our analysis, we therefore treat this cross-protocol misuse as a fundamental limitation of the baseline design whenever
$K_{\mathrm{master}}$ is derived directly from raw PKCS\#11 signatures on a fixed label. In some controlled deployments,
it may be possible to enforce, at the policy or middleware level, that the QES token is used exclusively for FIDO-related
operations, thereby preventing such misuse by non-WebAuthn applications. However, such enforcement is not universally
available, particularly for general-purpose PKCS\#11 tokens and heterogeneous client environments.

To address this limitation in a deployment-agnostic manner, we introduce in Section~\ref{sec:oprf-hardening} an optional
hardened architecture that augments key derivation from $\sigma$ with an Oblivious Pseudorandom Function (OPRF)-based
mechanism bound to a local user-verification factor. At a high level, the hardened variant ensures that the token never
exposes a reusable value from which $K_{\mathrm{master}}$ can be reconstructed in other protocol contexts, thereby closing
this cross-protocol attack channel while preserving the overall architecture.

\subsection{Other Limitations}

The proposed architecture inherits a well-known limitation of local WebAuthn authenticators.
Once the user has legitimately unlocked the VFA using the PKCS\#11 token and PIN, subsequent WebAuthn operations may be
performed without repeating user verification. Remote malware that executes \emph{after} this unlock and is able to invoke
the CTAP2 interface may therefore trigger WebAuthn assertions during the unlocked window, even though it cannot extract
private keys or continue using them once the VFA is locked again. While platform authenticators may mitigate this risk
through trusted execution environments or additional policy checks, preventing such post-unlock misuse in a general and
portable manner lies largely outside the current WebAuthn threat model.

In addition, the system does not address collusion scenarios in which a user willingly cooperates with a remote attacker (e.g., via screen sharing) to perform authentications, nor does it mitigate phishing attacks that fully control the client UI if the user approves token unlocks. Finally, the security guarantees depend on the correct implementation and certification level of the PKCS\#11 token; weaknesses in token firmware, side-channel resistance, or PIN policies would directly impact the strength of the master-key protection.

Overall, while the architecture significantly improves key-confidentiality guarantees under cloud compromise and device theft and maintains server-side simplicity, it does not eliminate all forms of client-side compromise. These trade-offs differ from those of hardware FIDO2 authenticators, which typically place even weaker trust assumptions on
the client by keeping private keys strictly within dedicated hardware. Instead, our approach occupies a middle ground
between hardware-bound authenticators and cloud-synchronized passkeys, offering improved portability while maintaining a
more tightly scoped trust model than platform-managed passkey ecosystems.

\section{OPRF Hardening}
\label{sec:oprf-hardening}

The cross-protocol attack described in Section~\ref{sec:cross-protocol} arises because the baseline architecture derives the master key $K_{\mathrm{master}}$ directly from a deterministic PKCS\#11 signature on a fixed label. This deterministic signature
\[
\sigma = \mathrm{Sign}_P\!\left(H(\text{label})\right)
\]
must remain stable across devices so that $K_{\mathrm{master}}$ can be consistently recovered. However, this stability also means that if an attacker succeeds in triggering \emph{any} external protocol to compute the same signature on $H(\text{label})$, they immediately obtain $\sigma$ and can derive $K_{\mathrm{master}}$ offline. The goal of the hardened architecture is therefore not to eliminate deterministic signing—which is necessary for portability—but to ensure that $\sigma$ alone is no longer sufficient to derive the master key.

\subsection{Oblivious Pseudorandom Functions (OPRFs)}
\label{sec:oprf-intro}

Oblivious Pseudorandom Functions (OPRFs) provide a cryptographic mechanism in which a client obtains the value of a pseudorandom function $F_k(x)$ on an input $x$, while a server holding the secret key $k$ learns nothing about $x$, and the client learns nothing about $k$ beyond the single evaluated value. OPRFs have been extensively studied and applied in the design of privacy-preserving key-derivation and authentication protocols~\cite{oprf,opaque}. Their defining property---combining client-side input privacy with server-side key privacy---makes them a natural tool for deriving secrets that must remain bound to a user-controlled input (e.g., a PIN) without exposing either the input or the PRF key to other parties.

In the context of authentication, OPRFs are often used to strengthen the derivation of long-term secrets from low-entropy or user-controlled inputs. For example, the OPAQUE protocol~\cite{opaque} leverages OPRFs to transform user passwords into hardened cryptographic artifacts that resist offline dictionary attacks. More generally, OPRF-based derivations ensure that even if an adversary learns the result of the PRF computation, they cannot apply it to other inputs or learn any information about the underlying PRF key.

These properties make OPRFs particularly well suited to our setting. As discussed in Section~\ref{sec:cross-protocol}, the baseline architecture derives the master key $K_{\mathrm{master}}$ from a deterministic PKCS\#11 signature that must remain stable across devices. This determinism, however, allows an attacker who can trigger the same signing operation in a different protocol context to reconstruct $K_{\mathrm{master}}$. Introducing an OPRF into the derivation chain enables us to bind $K_{\mathrm{master}}$ not only to the token-provided deterministic signature but also to a user-verification factor in a way that cannot be reproduced by external protocols. The result is a derivation that preserves portability and determinism while eliminating the cross-protocol attack vector.

\subsection{High-Level Construction}

To preserve determinism but eliminate direct recoverability, the hardened variant introduces an Oblivious Pseudorandom Function (OPRF) evaluation over a user-verification factor. Let $x$ be a PIN-derived input, and let $F_{k_{\mathrm{OPRF}}}$ denote the OPRF with key $k_{\mathrm{OPRF}}$. The hardened derivation proceeds as follows:

\begin{enumerate}
    \item The user enters a local verification secret $x$.
    \item The VFA runs an OPRF protocol on the input $x \parallel \text{label}$ to obtain
    \[
        y = F_{k_{\mathrm{OPRF}}}(x \parallel \text{label}),
    \]
    while the OPRF server learns nothing about $x$.
    \item The VFA obtains the deterministic token signature $\sigma = \mathrm{Sign}_P(H(\text{label}))$, identical to the baseline design.
    \item The master key is derived as
    \[
        K_{\mathrm{master}} = \mathrm{HKDF}\!\big(\sigma \parallel y, \text{``VFA-MK-OPRF''}\big).
    \]
\end{enumerate}

Critically, both $\sigma$ and $y$ are required to reconstruct $K_{\mathrm{master}}$.  
The deterministic signature still ensures device portability, but no longer acts as a stand-alone secret.

\subsection{Mitigation of Cross-Protocol Misuse}

Under this design, possession of the deterministic signature $\sigma$ is insufficient to recover $K_{\mathrm{master}}$. Even if a malicious application or compromised middleware induces the token to sign $H(\text{label})$, the adversary must also obtain the correct OPRF output $y$ corresponding to the user’s verification secret $x$.

Since:
\begin{itemize}
    \item the OPRF is unlinkable and hides $x$ from the server,
    \item $y$ is not stored or reused across contexts,
    \item and the OPRF input is bound to both $x$ and the VFA-specific label domain,
\end{itemize}
the attacker cannot derive $K_{\mathrm{master}}$ unless they can additionally run the OPRF protocol with the correct input. This eliminates the attack path where a single PKCS\#11 signing operation in an unrelated context would otherwise reveal the master key.

\subsection{Reducing Cloud Trust via OPRF-Based Key Derivation}
\label{sec:oprf-trust}

A conventional approach to deriving or protecting synchronizable WebAuthn credentials is to rely on a keyed key-derivation
function (KDF), potentially combined with a cloud-provided salt or auxiliary secret. While such constructions can offer strong
cryptographic guarantees, their security fundamentally depends on the cloud acting as a trusted participant in the derivation process.

A representative example is FeIDo~\cite{feido}, which derives WebAuthn credentials deterministically using a long-term secret
$s_{\mathrm{kdf}}$ combined with relying-party context and user attributes. In FeIDo, the confidentiality of
$s_{\mathrm{kdf}}$ is enforced by executing the KDF inside a Trusted Execution Environment (TEE) deployed in the cloud. As a
result, the security of the construction relies on strong platform trust assumptions: the cloud provider is assumed to
correctly instantiate and protect the TEE, and adversaries are assumed unable to extract or misuse $s_{\mathrm{kdf}}$ from
within the enclave.

In contrast, our OPRF-based hardening adopts a strictly weaker trust model. Rather than assuming a cloud-hosted secret remains
confidential, the OPRF construction ensures that the cloud never becomes a trusted secret holder in the first place. Even if
the OPRF service is actively malicious or fully compromised, it cannot evaluate the key-derivation function autonomously,
perform offline guessing attacks on user-verification inputs, or reconstruct the resulting master key without interacting
with the client for each protocol execution. The cloud is thus reduced to a rate-limited online oracle, rather than a trusted
derivation authority.

Importantly, the security benefit of the OPRF-based approach lies not in replacing the KDF, but in preventing the cloud from
supplying all inputs necessary to evaluate it independently. This shift from confidentiality-based trust (as required by
cloud-hosted KDF designs) to obliviousness-based trust represents a key advantage for high-assurance deployments where reliance
on TEEs or cloud-enforced secrecy is undesirable or operationally infeasible.

Table~\ref{tab:feido-comparison} contrasts FeIDo with the two design options considered in this work.
The baseline design places the weakest trust assumptions on the cloud, as the synchronization service stores only ciphertext
and does not hold any protocol or derivation secret. The OPRF-hardened variant intentionally introduces a limited
protocol-internal secret on the cloud side in order to mitigate cross-protocol misuse of deterministic token operations.
Importantly, this additional trust does not enable independent key derivation or offline recovery by the cloud, but it
represents a deliberate trade-off between minimizing cloud trust and strengthening misuse resistance.

\begin{table*}[t]
\centering
\caption{Comparison of FeIDo and the Two Design Options Considered in This Work}
\label{tab:feido-comparison}
\small
\setlength{\tabcolsep}{6pt}
\begin{tabular}{p{3.6cm} p{4.1cm} p{4.1cm} p{4.1cm}}
\toprule
\textbf{Aspect}
& \textbf{FeIDo~\cite{feido}}
& \textbf{Baseline (This Work)}
& \textbf{OPRF-Hardened (This Work)} \\
\midrule
WebAuthn compatibility
& Standard WebAuthn
& Standard WebAuthn
& Standard WebAuthn \\

Credential generation
& Deterministic derivation (KDF)
& Random generation + encrypted storage
& Random generation + encrypted storage \\

Cloud involvement
& Active derivation service
& Ciphertext-only storage
& Ciphertext-only storage (plus OPRF service) \\

Cloud-held secret enabling key derivation
& Yes ($s_{\mathrm{kdf}}$, protected by TEE)
& No
& No$^{\dagger}$ \\

Trusted Execution Environment (TEE)
& Required
& Not required
& Optional (for OPRF deployment) \\

User-verification factor
& eID possession
& Token PIN
& Token PIN + OPRF \\

Resistance to cloud compromise
& Relies on TEE security
& Cryptographically enforced
& Cryptographically enforced \\

Cross-protocol misuse resistance
& Not explicitly addressed
& Limited
& Explicitly addressed via OPRF \\

Trust assumptions on cloud
& Cloud trusted to protect TEE
& Cloud assumed passive
& Cloud may be malicious \\

Implemented in this work
& No
& Yes
& No (analyzed) \\
\bottomrule
\end{tabular}

\begin{minipage}{\textwidth}
\vspace{2pt}
\footnotesize
$^{\dagger}$The OPRF service holds a protocol-internal secret (the PRF key) but no cloud-resident secret is sufficient to
derive or recover the master key independently. Key derivation requires interactive participation of the client and cannot
be performed offline by the cloud.
\end{minipage}
\end{table*}

% =============================================================
\section{Conclusion and Future Work}

This paper explored architectural options for enabling secure, multi-device portability of FIDO2 credentials while
preserving a strict hardware-rooted trust model. We designed and implemented a baseline system in which a Virtual FIDO2
Authenticator (VFA) encrypts all credential material under a master key derived from a PKCS\#11 e-signature token, allowing
ciphertexts to be synchronized across devices via an untrusted cloud service. Our implementation demonstrates that the
approach is practical, imposes modest performance overhead, and remains fully compatible with existing WebAuthn
relying-party infrastructures without requiring any server-side modifications.

We further identified a class of cross-protocol misuse attacks inherent to deterministic token operations and introduced an
optional OPRF-based hardening mechanism that binds the derivation of the master key also to a local user-verification factor.
While we leave the hardened variant as an architectural analysis rather than an implemented prototype, our results show
that such a mechanism can close the identified attack vector without sacrificing portability or FIDO2 compatibility.

Several promising directions remain for future work. First, integrating qualified electronic signature (QES) metadata or
certificate chains into attestation could enable relying parties to reason about the identity assurance level of the VFA
itself. Second, evaluating token support for EdDSA, deterministic ECDSA variants, or emerging post-quantum signature
schemes would broaden the applicability of the architecture as cryptographic standards evolve. Third, formal analysis of
both the baseline and hardened constructions under established security models for WebAuthn, OPRFs, and PKCS\#11 APIs would
strengthen confidence in the security guarantees.

Beyond the client-side architectures considered in this paper, an alternative direction would be to store FIDO2
credential private keys within a remote hardware security module (HSM) or signing service, and to use a local QES token
solely to authenticate and authorize access to this remote signer. Such an approach is conceptually similar to remote
qualified electronic signature deployments, but it shifts the trust boundary toward a centralized signing infrastructure
and introduces additional considerations related to latency, availability, auditability, and governance. As our focus is
on client-side designs that treat the cloud as untrusted with respect to key usage and derivation, we leave the systematic
analysis of remote-signing-based FIDO architectures as an out-of-scope but interesting direction for future work.

Finally, standardizing a PKCS\#11-backed roaming authenticator profile—or incorporating similar ideas into future WebAuthn
extensions—may help unify hardware-rooted identity systems with cloud-assisted passkey ecosystems. Overall, our study
illustrates that high-assurance authentication and multi-device usability need not be mutually exclusive. By anchoring
credential protection in existing PKCS\#11 tokens and separating trust in hardware from trust in cloud synchronization,
the proposed architectures offer a viable path toward deployable, secure, and interoperable roaming FIDO2 credentials.

%\section*{Acknowledgment}

%This research was partially funded by the Scientific and Technological Research Council of Türkiye (TÜBİTAK) under grant number 3230252.

% =============================================================
\bibliographystyle{IEEEtran}
\bibliography{references}

\end{document}